\documentclass[twocolumn,showpacs,prl,amsmath,amssymb]{revtex4}

\usepackage{graphicx}
\usepackage{bm}
\usepackage{mathbbol}
\DeclareMathOperator{\Tr}{Tr}

\DeclareMathAlphabet{\mathpzc}{OT1}{pzc}{m}{it}
\begin{document}

\title{Multiple-pulse coherence enhancement of solid state spin qubits}
\author{W. M. \surname{Witzel}} 
\author{S. \surname{Das Sarma}}
\affiliation{Condensed Matter Theory Center,
  Department of Physics, University of Maryland, College Park, MD
  20742-4111} 
\date{\today}
\begin{abstract}
We describe how the spin coherence time of a localized electron spin
in solids, i.e. a solid state spin qubit, can be prolonged
by applying designed electron spin resonance pulse sequences.  In
particular, the spin echo decay due to the spectral diffusion of the
electron spin resonance frequency induced by the non-Markovian
temporal fluctuations of the nuclear spin flip-flop dynamics can be
strongly suppressed using multiple-pulse sequences akin to the 
Carr-Purcell-Meiboom-Gill pulse sequence in nuclear magnetic resonance.  
Spin coherence time can be enhanced by factors of
4-10 in GaAs quantum dot and Si:P quantum computer architectures
using composite sequences with {\it an even number} of pulses.
\end{abstract}


\maketitle
Understanding spin coherence in solids is one of the oldest problems
in condensed matter physics~\cite{abragam62, slichter}, going back to the
seminal pioneering work of Hahn~\cite{hahn50}, Feher~\cite{feher59}, and
others. Recently, there has been a resurgence of widespread interest
in this problem in the context of quantum computation using spin
qubits in semiconductors.  In particular, a necessary condition for
quantum computation is long qubit coherence time so that quantum error
correction can be meaningfully implemented.  For various proposed
electron spin qubits in semiconductor nanostructures, coherence times
on the order of $0.1~\mbox{ms}$ (or longer) are required as a
necessary condition; assuming typical gating
times of $100~\mbox{ps}$, this leads to a Q-factor of 
$\sim 0.1~\mbox{ms} / 100~\mbox{ps} = 10^{6}$ which satisfies the
current quantum error correction constraint ($\sim 10^{-4} - 10^{-6}$).
In fact, the {\it only} reason for the current
experimental and theoretical interest in spin-qubit based scalable
solid state quantum computer architectures (as opposed to charge
qubits, which are easier to manipulate and read out) is the expected
long spin coherence times ($\gtrsim~\mbox{$\mu$s}$) compared with
charge (i.e. orbital) coherence times ($\lesssim~\mbox{ns}$) in
solids, making it impossible (possible) for quantum error correction
schemes to work for solid state charge (spin) qubits.
Understanding all aspects of electron spin decoherence in solid state
nuclear spin environment is important so that effective strategies can
be developed to restore and/or enhance coherence by, for example,
efficient pulse engineering.

It is, therefore, highly desirable to enhance the
coherence time of spin qubits in semiconductor structures,
particularly since semiconductor-based spin quantum computer
architectures have considerable advantages in terms of scalability and
fabrication.  It has been known for a long time that various
refocusing techniques using electromagnetic pulses (e.g. Hahn spin
echo) substantially enhance spin coherence compared with the
free induction decay.
In this Letter we describe techniques that
substantially enhance coherence of solid state spin qubits by
utilizing various composite sequences going beyond the simple
Hahn spin echo.
Demonstrating sufficient spin coherence in such a way
is a necessary step in the development of spin-based 
fault tolerant quantum computing.
This Letter has direct implications for quantum spin
memory in semiconductors and offers potential for 
coherent quantum logic gate design.

A particular impetus for our theoretical study comes from the
beautiful recent experimental work by Petta et al.~\cite{petta05} who
measured Hahn spin echo decay of electron spin qubits in GaAs
coupled quantum dot architectures, finding a spin coherence time $T_2
\sim 1~\mbox{$\mu$s}$
(consistent with our spectral diffusion theory of spin
decoherence in GaAs quantum dot spin qubits \cite{witzelSDlong}), 
substantially enhancing (by a factor of $100$)
the inhomogeneous spin dephasing time of $T_2^* \sim 10~\mbox{ns}$ in
the same system.  
In the current Letter, we investigate the potential
for further coherence time enhancement by
utilizing a more complex pulse sequence originally used by Carr
and Purcell a long time ago~\cite{meiboom58} in the context of
magnetic resonance studies.  
Further coherence enhancement in GaAs quantum dots has
recently been achieved using the multiple pulse sequence suggested
in this Letter, and the experimental results are in good agreement 
with our predictions \cite{marcus}.
Earlier experimental work on the ensemble
of P donor electronic states in Si, of great interest to Si quantum
computer architectures, reported very long ($\gtrsim~\mbox{ms}$) 
spin echo $T_2$ coherence times~\cite{tyryshkin, abe}, 
which can be further enhanced in Si through isotopic
purification.  We find that composite sequences 
beyond the simple Hahn echo sequence could lead to substantial spin coherence
enhancement in Si spin qubits also.

We consider a single 
localized electron spin in a semiconductor (e.g., GaAs quantum dot,
Si:P) interacting with the surrounding nuclear spin bath.  
The electron's spin polarization is preserved by applying a
magnetic field due to the very large (a factor of $\sim 2000$)
difference between the electron and nuclear Zeeman energy splittings.
Virtual electron spin-flip transitions allow hyperfine-mediated
interactions between nuclei, but such processes cause only
a small visibility decay of refocused echoes that is
sufficiently small for fault tolerant quantum computing at fields above
$B \sim 1~\mbox{T}$ \cite{shenvi} in GaAs quantum dots for which the effect is
particularly strong.
Echo modulation due to anisotropic hyperfine coupling is also 
sufficiently suppressed 
at fields above $B \sim 9~\mbox{T}$ \cite{saikin} in Si:P for which the effect
is particularly strong.
In such a situation (i.e. an applied magnetic field and low
$\lesssim 100~\mbox{mK}$ operational temperatures), it is now well
accepted that the decoherence of a solid state electron spin qubit is
dominated by the spectral diffusion process, where the flip-flop
dynamics of interacting nuclear spins creates a temporally random
non-Markovian magnetic field at the electron spin location leading to
decoherence.  We have earlier theoretically studied spectral diffusion
effects on the Hahn spin echo introducing a quantum
cluster expansion technique \cite{witzelSDshort, witzelSDlong} that
provides a formally exact treatment of the 
non-Markovian quantum
nuclear dynamics in the large applied magnetic field limit.  
In this Letter we apply this
technique to multi-pulse echo sequences 
in order to investigate the possible enhancement
of spin coherence.  We find that carefully
designed multi-pulse sequences could considerably enhance
solid state spin qubit coherence with the spin decoherence time
$T_2$ \cite{foot1}
increasing strongly with the number of pulses.
Our work has obvious implications for the design and operation of spin
quantum computer architectures in semiconductor nanostructures.
In
particular, we find that our non-Markovian treatment leads to low
order symmetry related cancellations when an even number of pulses are
applied.
We also predict that the logarithm of the 
echo, as a function of inter-pulse time (i.e., $\tau$), 
scales as the square of the number of applied pulses in stark contrast
to linear scaling of the stochastic theory \cite{desousa05}.

Spins of interest (e.g., spins representing qubits) can be refocused,
partially reversing the effects of their local magnetic fields,
by applying to them sequences of rotating pulses (e.g., via resonance).
Our analysis treats these pulses as ideal, making the approximating
assumption that they perform exact rotations of some desired spin in an
infinitesimal time without affecting the rest of the system.
We consider the application of the Carr-Purcell-Meiboom-Gill (CPMG)
~\cite{meiboom58} pulse sequence 
(equivalent, for our purposes with ideal pulses, to the Carr-Purcell 
sequence)
to a solid state electron spin qubit.  
We represent this sequence as
$(\tau \rightarrow \pi \rightarrow \tau)^n$ by which we mean
that the system evolves freely for a time $\tau$,
we apply a $\pi$-rotation pulse
perpendicular to the applied magnetic field,
evolve the system freely for time $\tau$ again, and repeat this process $n$ times.  
The $n=1$ case is equivalent to the Hahn echo.  Except for comparison with the
Hahn echo, we treat the $n=2\nu$ case of an even number of pulses.  We
find analytically (within relevant approximations) that even pulses
are enhanced.
A very general cluster expansion technique was introduced in
Ref.~[\onlinecite{witzelSDshort}], where it was
successfully applied to Hahn echoes of Si:P and was 
recently verified in an independently developed theory \cite{yao05}.
We apply this technique now to the CPMG sequence.
This analysis begins by writing the exact expression for the echo of the
$(\tau \rightarrow \pi \rightarrow \tau)^{2\nu}$ sequence:
\begin{equation}
\label{CPMG_echo}
v_{\mbox{\tiny CPMG}}(\tau) = \frac{1}{M} \left|
\Tr{\{\left[U(2 \tau)\right]^{\nu} U(\tau) 
\left[U^{\dag}(2 \tau)\right]^{\nu} U^{\dag}(\tau)\}}\right|,~
\end{equation}
with $U(t) = \exp{\left(-\imath{\cal H}_{+} t\right)}
\exp{\left(-\imath{\cal H}_{-} t\right)}$, 
${\cal H}_{\pm} ={\cal H}_{B} \pm \sum_{n} A_{n} I_{nz} / 2$, and
${\cal H}_{B} \approx \sum_{n \ne m}' b_{nm}
I_{n+} I_{m-}  - \sum_{n \ne m} 2 b_{nm} I_{nz} I _{mz}$ where the
first summation is restricted to pairs of like nuclei (so that
Zeeman energy is preserved when they flip-flop).
The $\Tr{}$ operation in Eq.~(\ref{CPMG_echo}) 
traces over the states of the
nuclear bath and $M$ is the number of such states.
The $\{A_n\}$ are hyperfine coupling constants between the spins of the
electron and nucleus $n$; 
typically $\max{(A_n)} \sim 10^6~s^{-1}$ (with $\hbar = 1$ units).
Nucleus $n$ has spin operators denoted with $\bm{I}_n$ 
and its gyromagnetic constant is $\gamma_n$.
The $\{b_{nm}\}$ are dipolar coupling
constants between nuclei in the bath;
typically, $\max{(b_{nm})} \sim 10^2~s^{-1}$ so that $b_{nm} \ll A_n$
which is important for the convergence of our cluster expansion.
Details are
given in Ref.~[\onlinecite{witzelSDlong}].

Our cluster expansion is an approximation that is based upon the fact
 that we may exactly decompose Eq.~(\ref{CPMG_echo}) 
into a sum of all possible products of contributions from disjoint
sets of nuclei:
\begin{equation}
\label{exactClusterDecomp}
v_{\mbox{\tiny CPMG}}(\tau) =
\sum_{
\substack{
\left\{{\cal C}_i\right\}~\mbox{\scriptsize{disjoint}}, \\
{\cal C}_i \ne \emptyset,
}}
\prod_i v_{{\cal C}_i}'(\tau).
\end{equation}
The full proof and necessary conditions for such a decomposition 
is provided in Ref.~[\onlinecite{witzelSDlong}].
Each ${\cal C}_i$ denotes a set of nuclei and $v_{{\cal C}_i}'(\tau)$
is called the ``contribution'' from this set.  
A contribution has the property that it can only be significant 
when interactions between nuclei in the set are significant and 
no part is isolated from the rest.  
We consider only local dipolar interactions in the
current Letter, and thus contributions only arise when the nuclei in the
set are spatially clustered together; hence we refer to these sets as
clusters.  
Before we define the ``cluster contribution,'' $v_{{\cal
    C}}'(\tau)$, we first define $v_{{\cal C}}(\tau)$ as the
solution to Eq.~(\ref{CPMG_echo}) when only considering nuclei
contained in some set (or cluster) ${\cal C}$.  A cluster
contribution may then be recursively defined by and computed using
\begin{equation}
\label{vC'_recursive}
v_{\cal C}'(\tau) = v_{\cal C}(\tau) -
\sum_{
\substack{
\left\{{\cal C}_i\right\}~\mbox{\scriptsize{disjoint}}, \\
{\cal C}_i \ne \emptyset,~{\cal C}_i~\subset~{\cal C}
}}
\prod_i v_{{\cal C}_i}'(\tau),
\end{equation}
subtracting from $v_{\cal C}(\tau)$ the sum of all products
of contributions from disjoint sets of clusters contained in 
${\cal C}$.

The decomposition of Eq.~(\ref{exactClusterDecomp}) is useful when 
making an approximation that assumes cluster
contributions decrease with an increase in cluster size.  We can then
define the $k^{\mbox{\scriptsize th}}$ order of a cluster expansion by
\begin{equation}
\label{idealClusterExpansion}
v_{\mbox{\tiny CPMG}}^{(k)}(\tau) =
\sum_{
\substack{
\left\{{\cal C}_i\right\}~\mbox{\scriptsize{disjoint}}, \\
{\cal C}_i \ne \emptyset,~\lvert{\cal C}_i\rvert \leq k
}}
\prod_i v_{{\cal C}_i}'(\tau).
\end{equation}
Because of the difficulty of iterating through all possible disjoint
sets of clusters (below a given size), we make a further approximation
by relaxing the constraint that they be {\it disjoint} sets.
In this approximation, the logarithm of the 
$k^{\mbox{\scriptsize th}}$ order result
simply becomes the sum of all cluster contributions up to size $k$:
\begin{equation}
\label{clusterExpansion}
\ln{\left(v_{\mbox{\tiny CPMG}}^{(k)}(\tau)\right)} \approx
\sum_{\left|{\cal C}\right| \leq k}  v_{{\cal C}}'(\tau).
\end{equation}
This approximation may be tested (or corrections made) as described in 
Ref.~[\onlinecite{witzelSDlong}].
We wish to note the simplicity of implementing our approach using
Eqs.~(\ref{clusterExpansion}) and (\ref{vC'_recursive}) as compared to
a very recent diagrammatic approach \cite{saikin06}.

\begin{figure}
\includegraphics[width=3in]{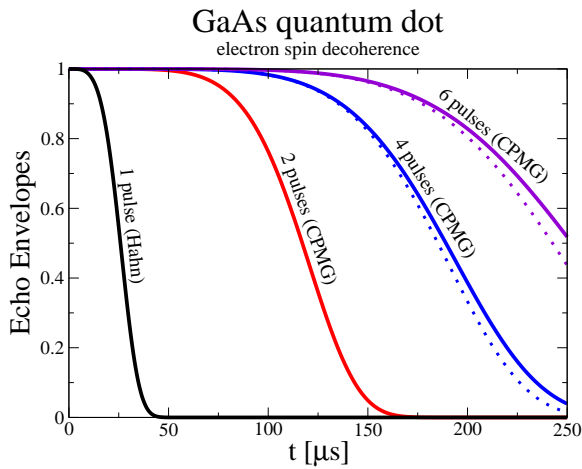}
\caption{
(color online). Numerical results for echo envelopes of the Hahn 
and various CPMG pulse
sequences (labelled by number of
pulses, $2 \nu$)
for a quantum dot in GaAs with a $50~\mbox{nm}$
Fock-Darwin radius, $10~\mbox{nm}$ quantum well thickness,
and applied magnetic field and thickness direction both along
$[001]$ in the lattice.  
Solid lines show convergent cluster expansion (exact) results while
dotted lines show lowest order perturbative results (within the
cluster expansion framework).
\label{figGaAs}}
\end{figure}

\begin{figure}
\includegraphics[width=3in]{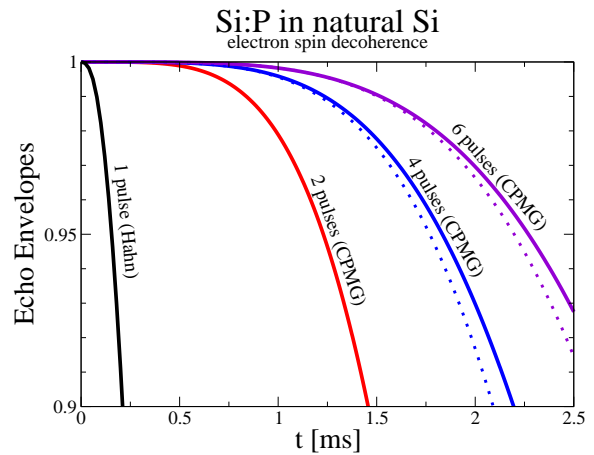}
\caption{
(color online). Numerical results for echo envelopes of the Hahn 
and various CPMG pulse
sequences (labelled by number of
pulses, $2 \nu$)
for the Si:P donor electron in natural Si 
with the applied magnetic field
along the $[001]$ lattice direction.
Solid lines show convergent cluster expansion (exact) results while 
dotted lines show lowest order perturbative results (within the
cluster expansion framework).  
The ordinate range is the first $10\%$ ($0.9$ to $1$) 
of echo decay where the cluster expansion is 
convergent for all pulse sequences shown.
\label{figSi}}
\end{figure}

We show numerical results obtained by this method (up to visible convergence) 
in Figs. \ref{figGaAs} and \ref{figSi} 
for the coherence of a quantum dot electron in GaAs and
for the coherence of an electron bound to a donor in Si:P
respectively, 
comparing Hahn and CPMG echo envelopes
Hahn echo envelopes.
The effect of the coherence enhancement for even echoes is 
noticeable in both
systems by
comparing results of two-pulse CPMG versus the Hahn echo.  
These figures also show that coherence is indeed prolonged
with an increase in the number of
pulses.  We note that these are plotted with respect to the total decay
time, $t = 4 \nu \tau$ (for the Hahn echo, $t = 2 \tau$).  

There are two perturbation theories that help to explain cluster
expansion convergence via diminishing cluster contributions with
increasing cluster size:
the ``inter-bath perturbation''
(previously called dipolar perturbation \cite{witzelSDlong}) that
treats the coupling between nuclei in the bath as a small perturbation
relative to the (potentially) strong coupling of the nuclei to the
electron, and the time (or $\tau$) expansion.  
Using $\lambda$ to generically denote the (small) perturbation
parameter of either theory, it was shown \cite{witzelSDlong} that a
contribution from a cluster of size $k$ is $O(\lambda^k)$.  The
inter-bath perturbation is most applicable to nuclei that are coupled
strongly to the electron (e.g., those closer to the center),
and the time expansion is most applicable to nuclei with a weak
coupling to the electron (e.g., those further from the nuclei where lower
energies result in slower evolution).  The two theories work in tandem
to produce overall convergence of the cluster expansion for a variety
of physical systems.

By using the lowest order results of cluster contributions 
[Eq.~(\ref{vC'_recursive})] with respect to either perturbation theory
in the computation of the cluster expansion
[Eq.~(\ref{clusterExpansion})], we can see the important role played
by these perturbation theories and gain useful insights.
The lowest order result with respect to inter-bath perturbation explains 
the enhancement of even CPMG echoes relative to odd
echoes as well as the scaling behavior with the number of pulses.
Using $\lambda \sim b_{nm} / \lvert A_n-A_m \rvert$ as the
perturbation parameter, the low order solution to Eq.~(\ref{CPMG_echo}) applied
to a given cluster, $v_{\cal C}(\tau)$, behaves as
$v_{{\cal C}}(\tau) = 1 - O(\nu^2 \lambda^4)$.  For a cluster
contribution [Eq.~(\ref{vC'_recursive})] we then have 
$v_{{\cal C}}'(\tau) = O(\nu^2 \lambda^4)$, and from the
approximate cluster expansion of Eq.~(\ref{clusterExpansion}), 
\begin{equation}
\label{v_CPMG_exp_perturb}
v_{\mbox{\tiny CPMG}}(\tau) = \exp{(-O(\nu^2 \lambda^4))}.
\end{equation}
The analogous expression for the Hahn echo has $O(\lambda^2)$; hence the
extra symmetry of the time sequence in even-pulsed CPMG echoes leads
to a cancellation of the lowest $\lambda^2$ order.  
For this reason, we predict an enhancement of even CPMG
echoes over odd echoes when this lowest order
result is valid (we find that it is generally valid as long as
the cluster expansion itself is convergent as we observed and
explained in our Hahn echo analysis \cite{witzelSDlong}).
Furthermore, Eq.~(\ref{v_CPMG_exp_perturb}) shows that the logarithm of
the CPMG echo as a function of $\tau$ (but not $t$) scales with the number
of puses squared.

Similarly, the lowest order with respect to the time expansion
gives $v_{{\cal C}}(\tau) = 1 - O(\tau^6)$ so that 
$v_{\mbox{\tiny CPMG}}(\tau) = \exp{(-O(\tau^6))}$ for the short time 
behavior of the CPMG echoes.
The analogous expression for the Hahn echo has $O(\tau^4)$ which is
cancelled [much like $O(\lambda^2)$ of the inter-bath perturbation]
by symmetry in the time sequence of even-pulse CPMG echoes.
This short time behavior is exhibited by the GaAs system but not the
Si:P system (the reason relates to the shape of their
electron wavefunctions and resulting cluster contribution statistics).


Applying the lowest order inter-bath perturbation represented by 
Eq.~(\ref{v_CPMG_exp_perturb}) to the GaAs quantum dot system of
Fig.~\ref{figGaAs} and fitting to these numerical results yields
$\ln{\left(v_{\mbox{\tiny CPMG}}(\tau)\right)} \approx
-\nu^2 \left(\tau / 55~\mbox{$\mu$s}\right)^{6} 
-\nu^2 \left(\tau / 31\mbox{$\mu$s}\right)^{6}$ where the first
(second) term result from clusters of size two (three).
Interestingly, $3$-cluster contributions dominate as a consequence of
the low order $\lambda^2$ cancellation.
This approximation agrees with the $O(\tau^6)$ short time behavior that we
anticipated.  It also agrees with the exact results shown in 
Fig.~\ref{figGaAs} except as shown by the dotted lines which deviate,
in a conservative way (predicting overly quick decoherence), 
from their corresponding exact results.

For the GaAs system of Fig.~\ref{figGaAs}, 
the two-pulse CPMG echo prediction of $T_2 = 120~\mbox{$\mu$s}$ 
\cite{foot1} is more than four times the Hahn echo $T_2 = 28~\mbox{$\mu$s}$,
demonstrating the enhancement of even echoes.
More generally, in the conservative perturbation approximation,
$\nu^2 \tau^6$ factor implies that $\tau$ effectively scales as 
$\nu^{-1/3}$; this means that the time between pulses, $2 \tau$, 
must be shortened as one increases the number of
pulses, $2 \nu$, in order to yield the same
degree of coherence.  However, since the total pulse sequence time,
$t$, is given by $t = 4 \nu \tau$, we have
$T_2 = \nu^{2/3} \times 120~\mbox{$\mu$s}$.   Therefore,
coherence is enhanced by applying multiple CPMG pulses, but one is
experimentally 
limited by the minimum time allowed between pulses which is in turn
limited by the time needed to apply each pulse.

Applying the lowest order inter-bath perturbation represented by 
Eq.~(\ref{v_CPMG_exp_perturb}) to the Si:P system of
Fig.~\ref{figSi} (with $B \parallel [100]$)
and fitting to these numerical results yields
$\ln{\left(v_{\mbox{\tiny CPMG}}(\tau)\right)} 
\approx
-\nu^2 f^2 \left(\tau / 230~\mbox{$\mu$s}\right)^{4.3} 
-\nu^2 f^3 \left(\tau /74~\mbox{$\mu$s}\right)^{4.3}$
where $f$ represents the fraction of Si that are the $^{29}$Si
isotope, the only isotope of Si with a net spin that can
contribute to the spectral diffusion.  
The $f^2$ ($f^3$) dependent term comes from
$2$-cluster ($3$-cluster) contributions.
Note that isotopic purification will make the most impact when
$3$-cluster contributions dominate over $2$-cluster contributions 
because of their relative scaling with $f$.  
These two contributions become comparable
when $f \approx 0.7\%$.  This may be an important consideration for 
a cost-benefit analysis of isotopic purification of Si.

For natural Si ($f = 4.67\%$),
$\ln{\left(v_{\mbox{\tiny CPMG}}(\tau)\right)} 
\approx -\nu^2 \left(\tau / 0.63~\mbox{ms}\right)^{4.3}$.
This agrees with the exact results shown in 
Fig.~\ref{figSi} except as shown by the dotted lines which deviate,
in a conservative way (predicting overly quick decoherence), 
from their corresponding exact results.
For the two-pulse CPMG
echo ($\nu = 1$) of natural Si, $T_2 = 2.5~\mbox{ms}$, more than four times
longer than the Hahn echo $T_2 = 0.6~\mbox{ms}$,
demonstrating, again, the coherence enhancement of
even echoes over odd echoes.
In general, the above $\nu^2 \tau^{4.3}$ factor 
implies that $\tau$ effective scales as $\nu^{-0.47}$; 
since $t = 4 \nu \tau$, $T_2$ scales as $\nu^{0.53}$.  
This is comparable to, but not quite as
good as, the $\nu^{0.67}$ scaling of $T_2$ for GaAs.  For the natural
Si system of Fig.~\ref{figSi}, 
$T_2 = \nu^{0.53} \times 2.5~\mbox{ms}$, in this conservative perturbation approximation.

We have developed a cluster expansion theory for calculating the
enhancement of spin coherence in the context of multiple-pulse spin
echoes in the limit of large applied magnetic fields.  
We find considerable enhancement of spin coherence in
semiconductor qubits when even CPMG pulses are employed;
$T_2$ increases four fold in both GaAs and Si:P 
systems in comparing 1- (Hahn) and 2-pulse sequences,
and increases with additional pulses.
Using a conservative approximation for these GaAs and Si:P systems,
$T_2$ scales with the number of pulses to the power of $0.67$ and
$0.53$ respectively.
We also report a cross-over in the scaling of
even CPMG echoes as a function of isotopic purification of Si in Si:P
marking a transition between $3$-cluster and $2$-cluster dominance.
This is a further consequence of the interesting symmetry related
cancellation responsible for our important result that 
even-numbered pulse sequences are subtantially more effective in 
extending/restoring spin coherence than odd sequences.

This work is supported by DTO-ARO and NSA-LPS.


\begin{thebibliography}{99}
\bibitem{foot1}
We operationally define $T_2$ as the total pulse sequence time 
at which the echo envelope, though not a simple exponential,
 decays to $1/e$.
\end{thebibliography}
\end{document}